\documentclass[prb,aps,epsf,twocolumn,showpacs,superscriptaddress]{revtex4}
\usepackage{graphicx}
\usepackage{epsfig}
\usepackage{latexsym}
\usepackage{amsmath}
\begin{document}
\newcommand{\beq}{\begin{equation}}
\newcommand{\beqr}{\begin{eqnarray}}
\newcommand{\eeqr}{\end{eqnarray}}
\newcommand{\eeq}{\end{equation}}
\newcommand{\s}{{\sigma}}
\newcommand{\e}{{\varepsilon}}
\newcommand{\bofp}{{\bB (\bp )}}
\newcommand{\om}{{\omega}}
\newcommand{\baro}{{\bar{\rho}_{{\mbox{\tiny FCB}}}(\bq )}}
\newcommand{\Om}{{\Omega}}
\newcommand{\D}{{\Delta}}
\newcommand{\half}{{{1 \over 2}}}
\newcommand{\de}{{\delta}}
\newcommand{\al}{{\alpha}}
\newcommand{\ga}{{\gamma}}
\newcommand{\La}{{\Lambda}}
\newcommand{\be}{{\beta}}
\newcommand{\psib}{{\bar{\psi}}}
\newcommand{\rb}{{\bar{\rho}}}
\newcommand{\phib}{{\bar{\phi}}}
\newcommand{\dt}{{\Delta}}
\newcommand{\dva}{{\frac{\vp\times\va}{2\pi}-\rb}}
\newcommand{\dvab}{{\frac{\vp\times(\va-\vb)}{4p\pi}}}
\newcommand{\w}{{\omega}}
\newcommand{\zh}{{\hat{z}}}
\newcommand{\qh}{{\hat{q}}}
\newcommand{\vA}{{\vec{A}}}
\newcommand{\va}{{\vec{a}}}
\newcommand{\vrr}{{\vec{r}}}
\newcommand{\vj}{{\vec{j}}}
\newcommand{\vE}{{\vec{E}}}
\newcommand{\vB}{{\vec{B}}}
\def\bA{{\mathbf {\cal A}}}
\def\bm{{\mathbf m}}
\def\bsig{{\mathbf \sigma}}
\newcommand{\etab}{\mbox{\boldmath $\eta $}}
\def\bB{{\mathbf {\cal B}}}
\def\bp{{\mathbf p}}
\def\be{{\mathbf e}}
\def\bI{{\mathbf I}}
\def\bn{{\mathbf n}}
\def\bM{{\mathbf M}}
\def\bG{{\mathbf G}}
\def\bq{{\mathbf q}}
\def\bp{{\mathbf p}}
\def\br{{\mathbf r}}
\def\bx{{\mathbf x}}
\def\bs{{\mathbf s}}
\def\bS{{\mathbf S}}
\def\bQ{{\mathbf Q}}
\def\bq{{\mathbf q}}
\def\bs{{\mathbf s}}
\def\bi{{\mathbf i}}
\def\bj{{\mathbf j}}
\def\bG{{\mathbf G}}
\def\bl{{\mathbf l}}
\def\bPi{{\mathbf \Pi}}
\def\bJ{{\mathbf J}}
\def\bR{{\mathbf R}}
\def\bz{{\mathbf z}}
\def\ba{{\mathbf a}}
\def\bk{{\mathbf k}}
\def\bK{{\mathbf K}}
\def\bP{{\mathbf P}}
\def\bg{{\mathbf g}}
\def\bX{{\mathbf X}}
\newcommand{\Psib}{\mbox{\boldmath $\Psi $}}
\newcommand{\thetab}{\mbox{\boldmath $\theta $}}
\newcommand{\sigmab}{\mbox{\boldmath $\sigma $}}
\newcommand{\nablab}{\mbox{\boldmath $\nabla $}}
\newcommand{\gammab}{\mbox{\boldmath $\gamma $}}
\newcommand{\vx}{{\vec{x}}}
\newcommand{\vq}{{\vec{q}}}
\newcommand{\vQ}{{\vec{Q}}}
\newcommand{\vd}{{\vec{d}}}
\newcommand{\vb}{{\vec{b}}}
\newcommand{\vp}{{\vec{\partial}}}
\newcommand{\p}{{\partial}}
\newcommand{\gr}{{\nabla}}
\newcommand{\ra}{{\rightarrow}}
\def\dd{d^{\dagger}}
\def\half{{1\over2}}
\def\fifth{{1\over5}}
\def\third{{1\over3}}
\def\twof{{2\over5}}
\def\threes{{3\over7}}
\def\rhob{{\bar \rho}}
\def\ua{\uparrow}
\def\da{\downarrow}
\def\eqa{\begin{eqnarray}}
\def\eea{\end{eqnarray}}
\def\jetp{{\it Sov. Phys. JETP\ }}
\parindent=4mm
\addtolength{\textheight}{0.9truecm}
\title{Hamiltonian theory of fractionally filled Chern bands \\
}
\author{Ganpathy Murthy }
\affiliation{  Department of Physics and
Astronomy, University of Kentucky, Lexington KY 40506-0055   }
\author{R. Shankar}
\affiliation{Department of Physics, Yale University, New Haven CT 06520}

\date{\today}
\begin{abstract}

There is convincing numerical evidence that fractional quantum Hall
(FQH)-like ground states arise in fractionally filled Chern bands
(FCB).  Here we show that the Hamiltonian theory of Composite Fermions
(CF) can be as useful in describing the FCB as it was in describing
the FQHE in the continuum.  We are able to introduce CFs into the FCB
problem even though there is no external magnetic field by following a
two-stage process.  First we construct an algebraically exact mapping
which expresses the electron density projected to the Chern band,
${\rho}_{{\mbox{\tiny FCB}}}$, as a sum of Girvin-MacDonald-Platzman
density operators, ${\rho}_{{\mbox{\tiny GMP}}}$, that obey the
Magnetic Translation Algebra. Next, following our Hamiltonian
treatment of the FQH problem, we rewrite the GMP operators in terms of
CF variables which reproduce the same algebra. This naturally produces
a unique Hartree-Fock ground state for the CFs, which can be used as a
springboard for computing gaps, response functions,
temperature-dependent phenomena, and the influence of disorder. We
give two concrete examples, one of which has no analog in the
continuum FQHE with $\nu= {1 \over 5}$ and $\sigma_{xy}={2\over
  5}$. Our approach can be easily extended to fractionally filled,
strongly interacting two-dimensional time-reversal-invariant
topological insulators.
\end{abstract}

\maketitle

\section{Introduction and strategy}
Models with no net magnetic flux but with a quantized Hall conductance
$\sigma_{xy}$ have been known since the work of Haldane\cite{haldane}
and Volovik \cite{volovik}. The breaking of time-reversal symmetry,
necessary for $\sigma_{xy}\ne0$, manifests itself as a nontrivial
Berry flux for the band, whose non-zero integral over the Brillouin
zone (BZ) gives the Chern number ${\cal C}$. The work of Thouless {\em
  at al} \cite{TKNN}, equates ${\cal C}$ to the dimensionless Hall
conductance of the filled band. We use a convention in which  $\sigma_{xy}=-{\cal C}$. 
 
While we focus on single Chern bands, the approach to be described
here readily applies to strongly interacting two dimensional
time-reversal invariant topological insulators (2DTI's) \cite{TI-reviews,levin} which can be
thought of as pairs of time reversed Chern bands.

A question that has recently attracted much attention is whether these
FCB's could also exhibit the FQHE at partial filling in the presence
of suitable interactions.  In such cases they are called fractional
Chern insulators, or FCIs. Optimal conditions call for a hierarchy of
scales, where the band gap $\Delta$, the interaction strength
$V_{ee}$, and the FCB bandwidth $W$ obey $\Delta\gg V_{ee}\gg
W$. There have been three fronts of attack.  Numerical efforts
have concentrated on ``flattening'' the FCB\cite{tang,sun,neupert},
and realized Laughlin-like states by exact
diagonalization\cite{sun,neupert,sheng,bern,sheng-wang,neupert2}. Most
recently, other principal FQH fractions such as $2/5$ and $3/7$ have
been seen as well\cite{bern2}. On the analytical front, Qi \cite{qi}
has constructed a basis in which known FQHE wavefunctions can be
transcribed into the FCB.  Recently Wu, Regnault and Bernevig \cite{lrb} have pointed out that if Qi's plan is to yield wavefunctions with substantial overlap with the exact functions, his proposal must be modified to exploit a residual gauge freedom that maximizes the overlap. Several studies have likewise been devoted
to the parton construction for FCIs\cite{parton, swingle, luran} in
which the electron is fractionalized into quarks, each of which is in
an Integer Quantum Hall state.

Our work was stimulated by the third approach due to Parameswaran {\em
  et al} \cite{sid} who examined the algebra of ${\rho}_{{\mbox{\tiny
      FCB}}}(\bq)$, the density operators projected into the FCB.
Recall that in the LLL problem the projected density is essentially
${\rho}_{{\mbox{\tiny GMP}}}(\bq )$, the Girvin-MacDonald-Platzman
\cite{GMP} operator, which obeys the algebra of magnetic
translations: 

\beq \left[ {\rho}_{{\mbox{\tiny GMP}}}(\bq ),
  {\rho}_{{\mbox{\tiny GMP}}}(\bq' )\right]= 2 i \sin \left[ {
    l^2\over 2 }{\bq \times \bq' }\right] {\rho}_{{\mbox{\tiny
      GMP}}}(\bq +\bq') .\label{gmp} \eeq where \beq l= { 1 \over
  \sqrt{e B_0}} 
\eeq
 
is the magnetic length associated with the
perpendicular external field $B_0$.  By contrast, the algebra of
${\rho}_{{\mbox{\tiny FCB}}}(\bq)$ does not even close, though in the
small $\bq,\ \bq'$ limit the commutator is proportional to $\bq \times \bq'$. The fundamental reason for the non-closure of the density algebra is the varying Chern density $\bB(\bp)$ in the Brillouin Zone. 
Parameswaran {\em et al} offer interesting ways to combat the varying
$\bB$, such as smoothing it out or replacing it by its
average.

Our approach, by contrast,   tackles  the varying Chern density from the outset. It  is based on two indisputable facts:

\begin{itemize}
 \item We are looking for the FQHE in the FCB problem.
  \item  Composite Fermions are very useful in describing the FQHE problem in the continuum \cite{jain}.       \end{itemize}

It is then reasonable to ask if CFs can be made play an equally
fruitful rule in the FCB problem. Our answer is an emphatic yes. We
employ the Hamiltonian approach \cite{rmp}, which provides an operator
realization of CFs. In the past this has allowed us to  compute not only gaps but
also correlation functions at non-zero $q,\omega$ and $T$ and
disorder. At $\nu =\half$ it yielded relaxation rates and
polarization as a function of $T$. We describe how these ideas can be imported to the FCB problem.

The Hamiltonian theory of CFs presumes the existence of a uniform
external magnetic field $B_0$. To those who say "Where did the
magnetic field come from?", we say "Where did it go when the $\nu =
\half$ state was described as a Fermi sea?"\cite{hlr}. We adopt the
pragmatic view that one must choose whichever mapping takes us closest
to the desired end product. In the present case, when we want to
describe FQH-like physics in a Chern band, LL-based constructs are a
natural platform from which to make the leap. Furthermore, as shown by
the recent work of Wu, Jain, and Sun\cite{yjs}, the Hofstadter problem
is adiabatically connected to the Chern band problem.

\subsection{Brief history of the Composite Fermion}
Let us begin by asking what we mean by the CF, since the term has many
connotations.

It all began with the realization that in two dimensions the
statistics of particles could be altered by a singular gauge
transformation of the wavefunctions that essentially attached flux
tubes to the particles\cite{anyons}. For fractions of the form $\nu =
{1 \over 2s+1}$, Zhang, Hansson and Kivelson \cite{zhk} converted the
electron to a composite boson by attaching $2s+1$ flux quanta in the
path-integral formulation of a Chern-Simons theory. They then
explained many of the FQHE effects in terms of Bose condensation.
Jain\cite{jain} then discovered that when $\nu ={p\over 2ps+1}$, one
could get excellent trial wavefunctions by converting an electron to
another (Composite) fermion by attaching $2s$ flux quanta.  Lopez and
Fradkin\cite{lopez} implemented this flux attachment for Jain
fractions in a Chern-Simons path integral and computed response
functions.  These flux-attached CFs live in the full fermionic Hilbert
space, have a bare mass, and carry the same charge as the electron,
i.e, $e^*=e$. They proved especially useful in the gapless state at
$\nu = \half$, analyzed in depth by Halperin, Lee and Read \cite{hlr}.

The mean-field wavefunction due to {\em flux attachment} for fractions of the form $\nu = {p\over 2p +1}$ is

\beq
\Psi_{}^{CS}=\prod_{i<j}{(z_i -z_j)^2 \over |z_i-z_j|^2}\chi_p(z,\bar{z})
\eeq

where $\chi_p$ stands for p-filled CF-LLs.  Jain's ansatz  

\beq
\Psi_{}^{Jain}= {\cal P}\prod_{i<j}(z_i -z_j)^2 \chi_p(z,\bar{z})
\eeq
 
is obtained by dropping $|z_i-z_j|^2$ and projecting the $\bar{z}$
in $\chi_p$ to the LLL using ${\cal P}: \bar{z} \to 2 l^2 \p/\p z$.  The
double zero in the analytic Jastrow factor describes the charge
deficit due to a double vortex that follows the electron, leading to a
CF that has $e^*=e(1 - {2p\over 2p +1})={e \over 2p +1}$. This is the
CF obtained by {\em vortex attachment}.

In the path integral approaches \cite{zhk,lopez} the change
from flux attachment to vortex attachment is achieved by considering
fluctuations about the mean-field, while in our earlier Hamiltonian
approach\cite{prl-ms} plasmon correlations {\em a la} Bohm-Pines were
responsible.

\subsection{Brief review of the Hamiltonian theory}

We work \cite{rmp} with CFs that live in the LLL from the beginning,
as did Read\cite{read}, and Pasquier and Haldane \cite{pasqier}. Our CFs
carry both the phase and charge deficit of a double zero in the FQHE
wavefunction \cite{laugh,jain}.  Their entire Hamiltonian is given by
the electron-electron interaction projected to the LLL.
  
We introduce CFs as follows. The FQH problem projected to the LLL is
defined by the Hamiltonian

\beq
 \bar{H}= \half \sum_{\bq} {\rho}_{{\mbox{\tiny LLL}}} (\bq) v_{ee}(\bq ) {\rho}_{{\mbox{\tiny LLL}}} (-\bq)
\eeq
 
where ${\rho}_{{\mbox{\tiny LLL}}} (\bq)$ is the electron density
projected to the LLL. In first quantization the full electron density is 

\beq \rho (\bq)=
\sum_j e^{i \bq \cdot \br_e}. 
\eeq 

The electron's position $\br_e$ may
be decomposed as \beq \br_e = \bR_e + \etab_e \eeq where the
electronic guiding center coordinate $\bR_e$ and cyclotron coordinate
$\etab_e$ obey
\begin{eqnarray}
 \left[ R_{ex}, R_{ey}\right] &=&   -il^2 \label{straddle1} \\
 \left[ \eta_{ex}, \eta_{ey}\right] &=&   il^2 \label{straddle2}\\
 \left[ \etab_e , \bR_e\right]&=&0.
\end{eqnarray}
Upon projecting to the LLL

\beq
{\rho}_{{\mbox{\tiny LLL}}} (\bq) = \sum_{j} \langle e^{i \bq \cdot \etab_{ej}}\rangle_{{\mbox{\tiny LLL}}} e^{i \bq \cdot \bR_{ej}}= e^{-q^2 l^2 /4} {\rho}_{{\mbox{\tiny GMP}}}(\bq ).
\eeq
 
where each term $e^{i \bq \cdot \bR_{ej}}$ in the sum obeys the GMP
algebra by itself thanks to Eq. \ref{straddle1}.  We shall use the
same symbol for the densities when we switch to second quantization.

So the Hamiltonian to solve is 
\beq \bar{H}= \half \sum_{\bq} {\rho}_{{\mbox{\tiny GMP}}}(\bq )
\bar{v}_{ee}(\bq) {\rho}_{{\mbox{\tiny GMP}}}(-\bq )\label{hlll} \eeq
where $\bar{v}_{ee}(\bq) = v_{ee}(\bq ) e^{-q^2l^2/2}$. The problem is
difficult because, with $\etab_e$ projected to the LLL, the electron
is described by just one canonical pair $\bR_e$ and not two. The LLL
projected electron has half the degrees of freedom of a regular
two-dimensional fermion. However, the biggest problem is that at
fractional filling there is no clear mean-field state.

We attack these problems as follows. First we introduce an auxiliary
pair of conjugate "vortex" guiding center coordinates $\bR_v$. They
are defined by their commutation relations: \beqr \left[ R_{vx},
  R_{vy}\right] &=& i{l^2\over c^2} \\ c^2 &=& {2p\over 2p+1} \eeqr
Evidently the vortex describes a particle whose charge $-{2p \over
  2p+1}$ in electronic units is exactly that of the two vortices in
the Jastrow factor. It too has just half the degrees of freedom of a
regular two-dimensional particle.

We want these auxiliary coordinates to commute with everything electronic i.e.,
\beq
\left[ \bR_{e}, \bR_{v}\right]=0.
\eeq

{\em The cornerstone of our approach is that we can accommodate both
  $\bR_e$ and $\bR_v$ and their algebra very neatly into the Hilbert
  space of a regular two-dimensional fermion, which is going to be our
  composite fermion. }  This fermion feels the reduced field $B^*$
seen by a charge $e^*$ object.  In terms of its guiding center and
cyclotron coordinates (which carry no subscripts like $e$ or $v$) that
obey
\begin{eqnarray}
\left[ \eta_x , \eta_y \right]&=& il^{*2} = i{l^2\over 1-c^2}\\ \left[
R_x , R_y \right]&=& -il^{*2}\\
\left[ \etab , \bR \right]&=&0
\end{eqnarray}
the algebra of the two conjugate pairs
$\bR_e$ and $\bR_v$  can be realized   as follows:

\begin{eqnarray}
\bR_e &=& \bR + \etab \ c \\ \bR_v &=& \bR +\etab /c.
\end{eqnarray}

This in turn permits the crucial  {\em CF substitution} 
\beq
{\rho}_{{\mbox{\tiny GMP}}}(\bq )=\sum_{j} \exp \left[ i \bq \cdot (\bR_j + c \etab_j )\right]
\label{cfsub}\eeq

in Eqn. \ref{hlll} for the projected Hamiltonian $\bar{H}$, which now
acts on a regular fermionic Hilbert space with two conjugate pairs per
particle. Since the CFs see exactly the right field to fill an integer
number of CF-LLs, a natural, gapped Hartee-Fock state emerges.  The
price we pay for obtaining a good mean-field starting point is that
our Hilbert space has unphysical degrees of freedom $\bR_v$. In order
to work in the physical sector the vortex coordinates need to be
constrained. Specifically, the vortex densities, $\rho_v(\bq)=e^{i\bq
  \cdot \bR_v}$, emerge as a gauge algebra.  The way to  handle this
gauge degree of freedom is described in our review\cite{rmp}. 

The numbers computed in this scheme at nonzero $\omega, q,T$ and
disorder agree with data at the 10-15 \% level \cite{rmp}.

How is this formalism, predicated on making the CF-substitution in
${\rho}_{{\mbox{\tiny GMP}}}$, to be applied to the Chern band problem
where the density of interest is ${\rho}_{{\mbox{\tiny FCB}}}$? The
key is to establish the following algebraically exact mapping:

  \beq
{\rho}_{{\mbox{\tiny FCB}}} (\bq) = \sum_{\bG} c(\bG,\bq ) {\rho}_{{\mbox{\tiny GMP}}}(\bq + \bG)\label{47}
\eeq

{\em where the coefficients $c(\bG,\bq )$ can be computed from the
  data on the original Chern band, essentially by Fourier
  transformation.} The CF-substitution can be then made in each
${\rho}_{{\mbox{\tiny GMP}}}(\bq + \bG)$. While an explicit
demonstration follows later, here is the gist of the argument. On an
$N \times N$ toroidal lattice the number of possible values for $\bp $
and $\bq$ are $N^2$ each. We will show that the ${\rho}_{{\mbox{\tiny
      GMP}}}(\bq+\bG)$ are linearly independent only for $N^2$ values
of $\bG$ for each $\bq$ restricted to the Brillouin Zone (BZ). These
$N^4$ linearly independent operators ${\rho}_{{\mbox{\tiny GMP}}}(\bq
+ \bG)$ form a complete basis for one-body operators, just like the
canonical basis $d^{\dag }(\bp_i ) d(\bp_j)\ \ \left[ i, j=
  1...N^2\right]$.

The outline of the paper is as follows. In Section II we will show
that our approach applies to a (Type I) Chern band with a variable Chern density $\bB (\bp)$
and coulomb interaction between electrons. This Chern band is obtained
by starting with two electronic LLs and applying a periodic potential
$V_{{\mbox{\tiny PP}}}$ which mixes the LLs and causes $\bB (\bp)$ to
vary.  The lower band, the Modified Lowest Landau Level, or MLLL is
our Chern band with ${\cal C}=-1$. We explicitly derive Eq. \ref{47}
(with the MLLL being the FCB), compute the coefficients $c(\bG , \bq)$
and calculate the band structure of CFs in the HF approximation.

In Section III we show how, given a {\em specific} lattice Chern band and an
interaction between electrons, one can introduce CFs {\em without any
  reference to LLs. } We choose as our Type II example the Lattice Dirac Model
(LDM):

\beq H (\bp) = \sigma_1 \sin p_x +\sigma_2 \sin p_y +\sigma_3 (M -
\cos p_x- \cos p_y)\label{ldh} \eeq 

with $M=1$, which lies in the regime with ${\cal C}=-1$.  

Models in
which LLs appear or do not appear explicitly are labeled as Class I
and Class II respectively.

In Section IV we ask what happens if we apply our approach to a band with ${\cal C}=0$, while 
in Section V we turn our attention to FQH-like states that owe their
very existence to the lattice potential. Flux attachment on a lattice
was first investigated in the context of anyonic
states\cite{fradkin,eliezer}, and analyzed in the FQH context by Kol
and Read\cite{kol-read}. In such states, due to the explicit breaking
of Galilean symmetry, the dimensionless Hall conductance need not be
equal to the filling factor. There is suggestive numerical evidence of
such states in a problem of hard-core bosons in an external magnetic
field\cite{sorensen,moller}. By virtue of our mapping of the FCB problems to
LLL problems, such states should exist in FCBs as well. We define and
solve an illustrative Class I model at $\nu=\fifth$ and demonstrate
the existence of such a state where $\nu= {1 \over 5}$ and $\sigma_{xy}={2\over 5}$.

Section VI presents  conclusions and  open
questions.

\section{Class I Models}

The goal of this section is to convince the reader that a nonconstant
$\bofp$ is no impediment to the CF substitution, and to flesh out the
key expansion Eq. \ref{47}. We begin with the construction of a
nontrivial Chern band with a non-constant $\bofp$.  
  
  \begin{figure}[h]
\begin{center}
\includegraphics[width=10cm]{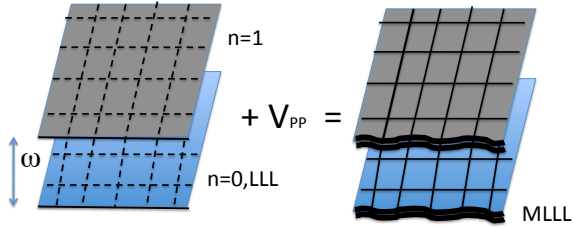}
\vspace*{-1in}
\caption{Left: Two unperturbed Landau Levels with ${\cal C}=-1$ and
  zero width separated by energy $\omega$. The dotted grid represents
  a fictitious square lattice with one flux quantum per unit
  cell. Right: The two bands with finite width after a periodic
  potential $V_{{\mbox{\tiny PP}}}$ (solid grid) is imposed. The lower
  of the two bands, the Modified Landau Level, the MLLL, is our Chern
  band with ${\cal C}=-1$. }
\label{mll} 
\end{center}	
\end{figure}
Consider a problem with two LLs labeled $0$ and $1$ separated by a gap
$\omega$ that is at our disposal, as shown in Fig.\ref{mll}. By
choosing the Hamiltonian to be a suitable function of
$\etab^{\dag}_{e}\etab_e$, not simply linear, we can arrange for the
other LLs to be separated by a parametrically larger gap than $\omega$
and hence ignorable in what follows.

Each level has ${\cal C}=-1$. It is instructive to demonstrate this
explicitly. (We recommend the review by Xiao, Chang and Niu for some
basic ideas of magnetic Bloch bands \cite{ming}.)  First we mentally
superpose on this continuum immersed in a perpendicular field $B_0$, a
square lattice of side $a$.  No real periodic potential is applied
yet.  Working in the Landau gauge \beq A_y(x,y)= xB_0
\ \ \ \ \ \ \ A_x(x,y)=0 \eeq we seek energy eigenfunctions which are
also simultaneous eigenfunctions of $T_x$ and $T_y$, the magnetic
translation operators in the $x$ and $y$ directions: \beq T_x=
e^{-iay_el^{-2}}e^{a \partial_x}\ \ \ \ T_y= e^{a \partial_y}.  \eeq
These commute with $H$, but not with each other unless each unit cell
has an integer number of flux quanta. We choose the simplest case of
one flux quantum penetrating each unit cell, i.e., \beq a^2= 2\pi l^2.
\eeq The simultaneous eigenfunctions we seek are\cite{TKNN}: \beqr
\lefteqn{\langle x_e,y_e|\bp,n\rangle = \Psi_{\bp, n} (x_e,y_e)}
\\ &=& \!\!\!\!{1 \over \sqrt{a}}\!\!\sum_{j=-\infty}^{\infty} \!\!
e^{iy_e(p_y+ a j l^{-2})}e^{ia p_xj}\phi_n(x_e\! - \! a j\!- \!
p_yl^2)\label{pn} \eeqr where $ \phi_n(x_e- a j- p_y l^2)$ is the
wavefunction for an oscillator in level $n$ centered at $x_e=a j+
l^2p_y$.

Hereafter we will set $a=1$  which means
\beq
l^2={1 \over 2\pi}.
\eeq
The states are normalized to unity, and   $\br_e$ integrals go over  the unit cell.

The Bloch functions are
\beqr
|u(\bp, n ) \rangle&=& e^{-i\bp \cdot \br_e} |\bp, n\rangle
\eeqr and the Berry connection

\beqr
\bA (\bp\ , n)&=& i \langle u(\bp,n) | \nabla_p|u (\bp ,n)\rangle
\eeqr

can be computed to have components

\beqr
{\cal A}_y &=&0\ \ \ \ \  {\cal A}_x= p_y l^2 = {1 \over 2 \pi} p_y\ \ \  \mbox{so that}\\
\bofp  &=& \nablab_p \times \bA = -{1 \over 2 \pi} \ \ \ \ \mbox{which means }\\
 {\cal C}&=&{1 \over 2 \pi}\int_{BZ}\bB=-1.
\eeqr

However this $\bB $ is constant in $\bp$ in both LLs. To make it vary,
we add a periodic potential 

\beq V(\br_e)= \sum_{\bG} V(\bG) e^{i
  \bG\cdot \br_e} \eeq 

which mixes the LLs and induces structure in
$\bofp$.  In our illustrative example we keep only the harmonics $\pm
2\pi$ in the two directions with coefficient $V_{10}$, though the
following analysis applies to the general case. Using 

\beq \langle \bp
n_2|e^{i \bG \cdot \br}|\bp n_1\rangle = \rho_{n_2n_1}e^{i
  G_xG_y/4\pi}e^{i \bG \times \bp /2\pi} 
\eeq 

where (for the general
value of $l$),

\beqr \lefteqn{\rho_{n_2n_1}(\bq )=
  e^{-q^2l^2/4}\sqrt{n_<!\over n_>!}L^{|n_1-n_2|}_{n_<}\left[{q^2l^2
      \over 2}\right]\nonumber }\\ &\times & \begin{cases}\left({i l
    \bar{z}\over \sqrt{2}} \right)^{|n_1-n_2|} \ \ \mbox{ when\ \ }
  n_1>n_2\\ \left({i l {z}\over \sqrt{2}} \right)^{|n_1-n_2|}
  \ \ \mbox{when}\ \ n_2\ge n_1\end{cases}\nonumber\\ z &=& q_x + i
  q_y\label{rhoq} \eeqr we find \beqr \lefteqn{H^{I}(\bp )= \left[
       g(\bp) -{\tilde{ V}} \sigma_1 \sin p_y +{\tilde{
          V}} \sigma_2 \sin p_x \right.\nonumber }
    \\ &-&\left. \sigma_3 \left[{\omega \over 2 } -
      {\sqrt{\pi}\tilde{V}\over 2}(\cos p_x + \cos
      p_y)\right]\right]\label{hp}\\ \tilde{V}&=& V_{10}e^{-\half \pi}
  \sqrt{\pi} \eeqr 

The ground state of $H^{I}(\bp)$ will be referred to as the Modified LLL,
or MLLL. It is our Chern band. The function $g(\bp)$ affects the
energy dispersion of the MLLL, but not the Chern density $\bofp$.

\begin{figure}[h]
\begin{center}
\includegraphics[width=8cm]{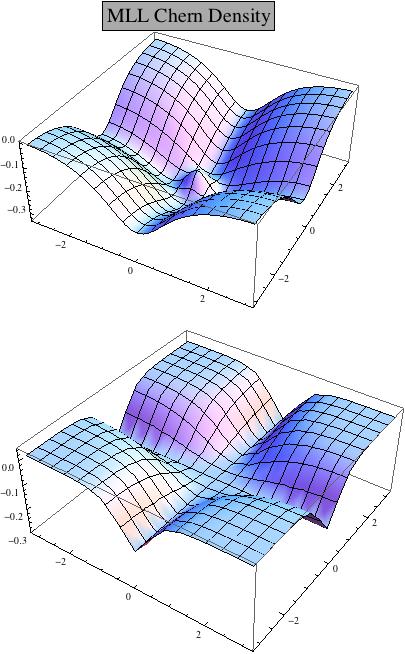}
\caption{Top:   $ \bofp $ in the MLLL  with just two LLs and one harmonic $V_{01}\ne 0$. Bottom: Berry flux density in the Lattice Dirac Model at $M=1$ . }
\label{actual} 
\end{center}	
\end{figure}

Though $H^{I}(\bp)$ has the form of the Lattice Dirac Model
(Eq. \ref{ldh}), it is in the topologically trivial region. This is
due to our requirement $\omega >2 \sqrt{\pi}\tilde{V}$ which ensures
that the two bands do not touch at $p_x=p_y=0$, which in turn ensures
that the Chern number remains ${\cal C}=-1$. Due to the topological
triviality of $H^{I}(\bp)$ the pseudo-spin $\bn (\bp) =\langle u(\bp) |
\sigmab |u(\bp) \rangle$ never enters the southern
hemisphere. Nonetheless the overall ${\cal C}=-1$ because {\em
  nontrivial topology is contained in the $\bp$-dependent basis
  functions.}  Whereas in the traditional LDM, the tight binding
wavefunctions are $\bp$-independent spinors, $\left[ 1,0\right]^T
\ \mbox{and}\ \left[ 0,1\right]^T$, here they are the states $|\bp ,
n=0 ,1 \rangle$ with topologically nontrivial $\bp$ dependence.  The
total $\bB$ in this problem has a constant piece $- {1 \over 2 \pi}$
coming from the basis functions and responsible for ${\cal C}=-1$, and
two more $\bp$-dependent terms with zero integrals: one due to the
$\bp$-dependence of the ground state spinor, and a cross term that
arises because $\langle n| \nablab_p |n'\rangle \ne 0$ for $n\ne n'$.
The total $\bB (\bp)$ is shown in Figure \ref{actual} along with the
Chern density for the LDM at $M=1$. Notice the strong similarity even
in this minimal model with just two LLs and one harmonic in $V$.

Now that we have a nontrivial Chern band with a nonconstant $\bofp$
let us proceed to the CF-substitution, which will in turn lead us to
the gapped state in the HF approximation at $\nu=\third$ when
interactions are turned on.

First we need to find $\rho_{{\mbox{\tiny MLLL}}}$, the projection of the
electron density operator to the Chern band, the MLLL. When
$V_{10}=0$, clearly $\rho_{{\mbox{\tiny MLLL}}} (\bq )=
\rho_{\mbox{\tiny LLL}} (\bq )$. To find out what it is when $V_{10}$
is turned on we proceed as follows:
\begin{itemize}
\item 
Find the $2 \times 2$ matrix that describes the electron density $\rho_e (\bq)=e^{i \bq \cdot \br_e }$ in the space of the LLs, $n=0,1$.
\item Find the eigenstates of $H^{I}(\bp)$.
\item Project $\rho_e (\bq)=e^{i \bq \cdot \br_e }$ to the ground state at each $\bp$, the MLLL.
\end{itemize} 

The matrix elements of $\rho_e (\bq)=e^{i \bq \cdot \br_e }$ between
the magnetic Bloch states defined in Eq. \ref{pn} vanish unless the
initial momentum $\bp$ and final momentum $\bp'$, both restricted to
the BZ, obey

 \beq
 \bp'=
 \left[ \bp + \bq \right]= \left( \bp + \bq \right) \ \ \ \mbox{mod \  $\bG$ }
 \eeq

Thus we must subtract from $\bp + \bq$ that $\bG$ which restricts
$\bp'$ to the BZ.  

\beq
2\pi N_x\be_x+2\pi  N_y \be_y=\bp+\bq-\left[\bp +\bq \right]
\eeq

The non-zero matrix elements are found to be

\beqr
\lefteqn{\langle \left[\bp + \bq\right] n_2|e^{i \bq \cdot \br_e   }|\bp n_1\rangle = \nonumber} \\& & \!\!\!\!\!\!\!\rho_{n_2n_1}\exp \left[ {i \over 2 \pi}\left(\half  q_xq_y+q_xp_y-(p_x+q_x)2 \pi N_y\right)\right]\nonumber \\
&\equiv&\rho_{n_2n_1}(\bq) e^{i\Phi(\bq,\ \bp)}\label{39}
\eeqr

where $\rho_{n_2n_1}(\bq)$ has been defined in Eq. \ref{rhoq}.
The asymmetry between $p_x$ and $p_y$ in Eq. \ref{39} reflects our
choice of the Landau gauge in defining the basis states in Eq. \ref{pn}.

In view of its importance to what follows we display
$e^{i\Phi(\bq,\ \bp)}$ prominently below: 

\beq
\boxed{e^{i\Phi(\bq,\ \bp)}=\exp \left[ {i \over 2 \pi}\left(\half
    q_xq_y+q_xp_y-(p_x+q_x)2 \pi N_y\right)\right]\label{etoiphi}}
\eeq

The corresponding second-quantized operator is
\beq
\rho_e(\bq)=
\sum_{\bp}\sum_{n_1,n_2=0,1}a^{\dag}_{n_2}( \left[ \bp+\bq\right]) \rho_{n_2n_1} (\bq) a_{n_1}(\bp)e^{i\Phi(\bq,\ \bp)}
\eeq

where $a_n$ and $a^{\dag}_n$ are the operators associated with the
basis states $|\bp n\rangle$.

Let $U$ be the matrix that diagonalizes $H^{I}(\bp)$ in Eq. \ref{hp} and  relates $a_n$ to the $d_n$ associated with the  energy eigenstates as follows

\beq
\left(
  \begin{array}{c}
    a_0 (\bp) \\
    a_{1}(\bp) \\
  \end{array}
\right)=\left(
          \begin{array}{cc}
            U_{00} & U_{01} \\
            U_{10} & U_{11} \\
          \end{array}
        \right)
\left(
  \begin{array}{c}
    d_0 (\bp) \\
    d_{1}(\bp) \\
  \end{array}
\right)
\eeq

Since $H^{I}(\bp)$ is topologically trivial, $U$, like the eigen-spinors,
is globally defined in the BZ and periodic in $\bp$. Switching to the
new basis and projecting to the ground state we obtain

\beqr
\rho_{{\mbox {\tiny MLLL}}}(\bq)&=&\sum_{\bp} d^{\dag}_{0}(\bp')d_0(\bp )e^{i\Phi(\bq,\ \bp)}f(\bq, \bp)\nonumber \\
f(\bq \ , \bp)&=& U^{\dag}_{0n'} (\bp')\rho_{n'n} (\bq) U_{n0}(\bp)\label{57}
\eeqr

Hereafter the subscript on $d_0$, indicating that it corresponds to
the ground state or MLLL, will be dropped.

Thus we have a Chern band, a non-constant $\bB$ and a closed
expression for the projected density.  The final step before we carry
out the CF substitution is to write this density in terms of
${\rho}_{{\mbox{\tiny GMP}}}$. Before doing this explicitly, we
provide an intuitive argument that this can be done.  When $V_{10}=0$,
we know $\rho_{{\mbox {\tiny MLLL}}}=e^{-q^2 l^2/4} \rho_{{\mbox
    {\tiny GMP}}}.$ As we turn on $V_{10}$, the perturbing terms are
of the form $e^{i \bG_0\cdot \br_e}=e^{i \bG_0\cdot \etab_e}e^{i
  \bG_0\cdot \bR_e}$ where $\bG_0= 2 \pi (\be_x n_x + \be_y n_y)$ with
only one of $n_x \ \mbox{or} \ n_y = \pm1$. Given the GMP algebra, the
repeated action of this perturbation can only be to turn $\rho_{{\mbox
    {\tiny GMP}}}(\bq) $ into a sum over $\rho_{{\mbox {\tiny GMP}}}(\bq
+ \bG)$, where $\bG$ is now any reciprocal lattice vector.  So we do
expect that in the end, even for an arbitrary periodic potential

\beq
{\rho}_{{\mbox {\tiny MLLL}}} (\bq) = \sum_{\bG} c(\bG ,\bq) {\rho}_{{\mbox{\tiny GMP}}}(\bq + \bG)\label{59}
\eeq

Since the bands never touch, perturbation theory will always
converge. However the final result is non-perturbative and follows
simply from the dependence of $H^{I}(\bp) $ on $e^{i \bG_0\cdot \bR_e} $.

We will now show Eq. \ref{59}  explicitly and compute $c(\bG, \bq)$.

First let us construct an auxiliary  operator which obeys the GMP algebra.

\beq \rho_{{\mbox {\tiny GMP}}} (\bq) =
\sum_{\bp}d^{\dag}_{}(\bp')d(\bp )e^{i\Phi(\bq,\ \bp)}\label{phigmp}
\eeq 

Given any BZ in which the operators $d,\ d^{\dagger}$ appearing in
Eq. \ref{phigmp} are canonical, it is easily verified that this
operator satisfies the magnetic translation algebra Eq. \ref{gmp}.
We likewise construct operators with momenta

\beq
\bq+ \bG =\bq+ 2\pi n_x \be_x+ 2 \pi n_y \be_y:
\eeq

defined by 

\beqr
\lefteqn{\rho_{{\mbox {\tiny GMP}}} ( \bq +\bG)\nonumber  }\\ &=& \sum_{\bp}d^{\dag}_{}(\bp')d(\bp )e^{i\Phi(\bq + \bG,\ \bp)}\nonumber\\
\!\!\!&=&\!\!\!\sum_{\bp}\!d^{\dag}_{}(\bp')d(\bp )e^{i\Phi(\bq ,\ \bp)}e^{{i \over 2}(q_yn_x-q_xn_y+2\pi n_xn_y)}e^{-ip_xn_y+ip_yn_x}\nonumber\\
&& \label{rhoq+g}
\eeqr

Note that whether we transfer momentum $\bq$ or $\bq+\bG$ to $\bp$ the
resultant $\bp'$ is the same. {\em We emphasize that these operators
can be constructed for any problem in a square lattice BZ, with no
reference to any LLs}. This fact will be crucial in the next section.

We now give the details of the counting argument that assures us that
$\rho_{{\mbox {\tiny MLLL}}} (\bq) $ may be expanded in terms of
$\rho_{{\mbox {\tiny GMP}}}(\bq + \bG)$. Consider a system of size
$L\times L$ wrapped into a torus. The question to ask is: For a given
$\bq$ in the BZ, for how many different values of $\bG$ are the
$\rho_{{\mbox {\tiny GMP}}}(\bq + \bG)$ linearly independent? Since
$a=1$, the number of sites is $N^2={L^2\over a^2} =L^2$, which also
equals the number of points in the $BZ$, the number of distinct values
for $\bp$ and the number of distinct values of $\bq$ in the lattice
model. The smallest value for any component of $\bq$ or $\bp$ is
$q_{min}= p_{min}={2\pi\over L}$. To verify that the largest distinct
value for any component of $\bG$ is $G_{max}= 2\pi N$, consider the
second and third exponentials in Eq. \ref{rhoq+g} which alone depend
on $\bG$. Focus on a factor like $e^{-{i\over 2}q_xn_y}$ when $q_x=
q_{min}$ and $n_y= N$ \beq \left. e^{-{i\over
    2}q_xn_y}\right|_{q_x={2\pi\over L}, n_y= N}=e^{-i\pi {N\over
    L}}=- 1.  \eeq The same goes for all the terms in the second
exponential, while the third exponential always equals unity, which
means \beq \rho_{{\mbox {\tiny GMP}}} ( \bq +\bG_{\mbox {\tiny max}})
\propto \rho_{{\mbox {\tiny GMP}}} ( \bq ) \eeq

Thus we get linearly independent densities only for components up to
$G_{max}= 2\pi N$. There are only $N^2$ independent values of $\bG$,
just as for $\bp$ or $\bq$.  But this means there are $N^4$ linearly
independent operators of the form $\rho_{{\mbox {\tiny GMP}}} ( \bq
+\bG)$, exactly the right number to form a basis, like the canonical
basis $d^{\dag}_{\bp_2}d_{\bp_1}$.  So what we find is that not only
$\rho_{{\mbox {\tiny FCB}}}$, but any bilinear operator $O$ (such as
the lattice current operator) of the form
$\sum_{\bp}d^{\dag}(\bp')d(\bp)O(\bq, \bp)$ can be expanded in terms
of $\rho_{{\mbox {\tiny GMP}}} ( \bq +\bG)$.

Having hammered home our central point, we now turn to the
determination of the coefficients of the expansion.  To this end we
combine Eqs. \ref{57}, \ref{59} and   \ref{rhoq+g} and write

\beqr\lefteqn{\rho_{{\mbox {\tiny MLLL}}}(\bq)\nonumber }\\&=&\sum_{\bp} d^{\dag}_{}(\bp')d(\bp )e^{i\Phi(\bq,\ \bp)}f(\bq, \bp)\label{romlll}\\
&=&\sum_{\bp}\sum_{n_x,n_y}c(n_x,n_y,\bq)d^{\dag}_{}(\bp')d(\bp )e^{i\Phi(\bq ,\ \bp)}\nonumber \\ &\times & e^{{i \over 2}(-q_xn_y+q_yn_x+2\pi n_xn_y)}e^{-ip_xn_y+ip_yn_x}\label{rogmp}
\eeqr

This equation can of course be satisfied since

\beqr
\lefteqn{f(\bq, \bp)=\sum_{n_x,n_y}c(n_x,n_y,\bq) }\nonumber \\ &\times & e^{{i \over 2}(-q_xn_y+q_yn_x+2\pi n_xn_y)} e^{-ip_xn_y+ip_yn_x}\label{idontcare}
\eeqr

is, at each $\bq$, just the Fourier expansion of the function $f$
periodic in $\bp$ in terms of oscillating exponentials of the right
period.

The commutators of the projected electron density $\rho_{{\mbox {\tiny
      MLLL}}}(\bq)$ can be worked out, if desired.  They will be
neither pretty nor universal\cite{sid}, unlike the magnetic
translation algebra\cite{GMP}, depending instead on the details of the
lattice via $f(\bq , \bp)$.

Having expressed $\rho_{{\mbox {\tiny MLLL}}}(\bq) $ in terms of
$\rho_{{\mbox {\tiny GMP}}}(\bq + \bG)$ we need to do the same for a
term in $\bar{H}$ that is absent in the usual LLL: the non-constant
kinetic energy $-\e (\bp)$ of the MLLL.  Because $\e(\bp )$ is periodic
mod $\bG$, this is a special case ($\bq =0 $) of the Fourier transform
we carried out for $\rho_{{\mbox {\tiny MLLL}}}(\bq) $. We write

\beqr
\lefteqn{-\sum_{\bp} d^{\dag}(\bp) d(\bp) \e (\bp) \nonumber }\\&=& \sum_{G} h(\bG ) \rho_{{\mbox {\tiny GMP}}}(\bG)\\
&=& \sum_{n_xn_y\bp}\ \!\!\!\!d^{\dag}(\bp) d(\bp)h(n_x,n_y )e^{-ip_xn_y+ip_yn_x+i\pi n_xn_y}\label{expansion}
\eeqr

which amounts to Fourier expanding the energy dispersion $-\e
(\bp)$. We now have the full electronic hamiltonian for the FCB in
terms of $\rho_{{\mbox {\tiny GMP}}}$s.

Note that the phase factor

\beq
e^{i\Phi(\bq,\ \bp)}=\exp \left[ {i \over 2 \pi}\left(\half  q_xq_y+q_xp_y-(p_x+q_x)2 \pi N_y\right)\right]\label{etoiphi2}
\eeq

jumps in $\bp$ space for a fixed $\bq$. For example if $\bq= 3\be_y $
and the BZ is in the interval $\left[ 0-2\pi\right]$ in both
directions, then for any $\bp$ with $p_y>2\pi -3$, adding $\bq$ will
take it to the next BZ and $N_y$ will have to jump from $0\ $ to $\ 1$
to bring $\bp'$ within the BZ. {\em Luckily, this discontinuous $\Phi$
  and its jump are shared by both $\rho_{{\mbox {\tiny MLLL}}}(\bq) $
  and $\rho_{{\mbox {\tiny GMP}}}(\bq +\bG) $} as we find in
Eqs. \ref{romlll} and \ref{rogmp}.  This ensures rapid convergence of
the Fourier expansion of the jump-free part $f(\bq ,\bp)$ in
Eq. \ref{idontcare}.

\subsection{The CF substitution}

Now we must switch to CFs. We only sketch the broad ideas.  We
consider the case of $\nu = {1 \over 3}$ when the CF has a charge
$e^*={1 \over 3}e$ and $l^{*2}= 3 l^2={3\over 2\pi }$.  The spatial
unit cell is 3 units long in the x-direction so as to enclose unit
flux {\em as seen by the CF} so that $BZ_{{\mbox {\tiny CF }}}$ goes
from $-{\pi \over 3}\le p_x\le {\pi \over 3}$ and is unchanged in the
y-direction. However when we construct the projected electron density
operators we will need to consider $\bq$ that runs over the BZ of the
electron not the CF.

Consider $\rho_{{\mbox {\tiny GMP}}}(\bq)$ which was $e^{i\bq \cdot
  \bR_e}$ in first quantization and

\beqr
\lefteqn{\rho_{{\mbox {\tiny GMP}}}(\bq)\nonumber}\\ &=&\sum_{\bp \in BZ_e}d^{\dag}(\bp')d(\bp)\nonumber \\ &\times&\exp \left[ {i \over 2 \pi}\left(\half  q_xq_y+q_xp_y-(p_x+q_x)2 \pi N_y\right)\right]
\eeqr

in second quantization. To go to the CF representation  means to write

\beq
e^{i\bq \cdot \bR_e}=e^{i\bq \cdot (\bR+\etab c)}
 \eeq
 
in first quantization and the following representation in terms of CF
operators $C$ and $C^{\dag}$ in second quantization

\beqr
\lefteqn{\rho_{{\mbox {\tiny GMP}}}(\bq)\nonumber}\\ &=&\sum_{\bp \in BZ_{CF}}C^{\dag}_{n'}(\bp')C_{n}(\bp)\rho_{n'n}( \bq \to c\bq, l \to l^*)\nonumber \\ &\times& \exp \left[ {3i \over 2 \pi}\left(\half  q_xq_y+q_xp_y-(p_x+q_x)2 \pi N_y\right)\right]
\eeqr

where the $3$ is due to $l^{*2}= 3 l^2={3\over 2\pi}$, and the
argument of $\rho_{n'n}$ is $\bq c$ because the $c$ in $\etab c$ may
be lumped with $\bq$ (see Eq. \ref{cfsub}). Note that all CF-LLs
($n=0,1,\dots$) appear in the density, a result of the enlarged
Hilbert space in which we are representing the problem.

\begin{figure}[t]
\includegraphics[width=8cm]{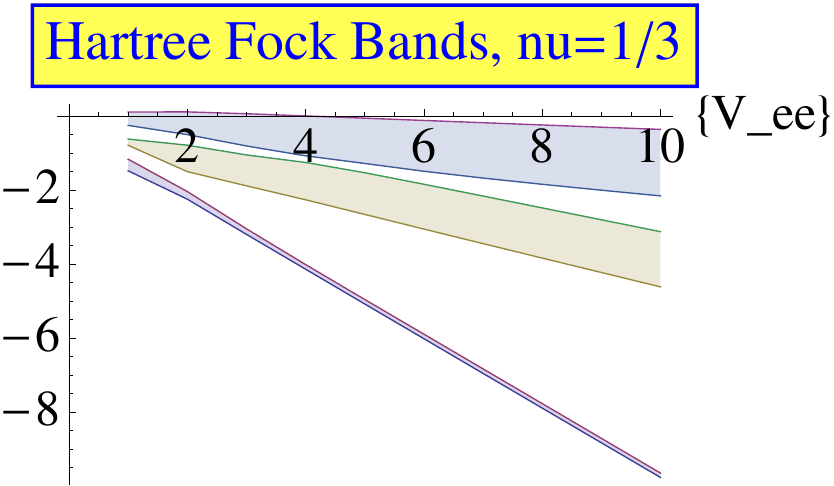}
\vspace*{0in}
\caption{ The results of our Hartree Fock calculation at $\nu ={1 \over 3}$ and $\omega=10$, for the Coulomb interaction ${2\pi V_{ee} \over q}$.  The three bands resulting from three CF-LLs which get mixed and modulated by the periodic potential $V_{{\tiny PPP}}=1$.}
\label{hf}
\end{figure}

\begin{figure}[h]
\includegraphics[width=10cm]{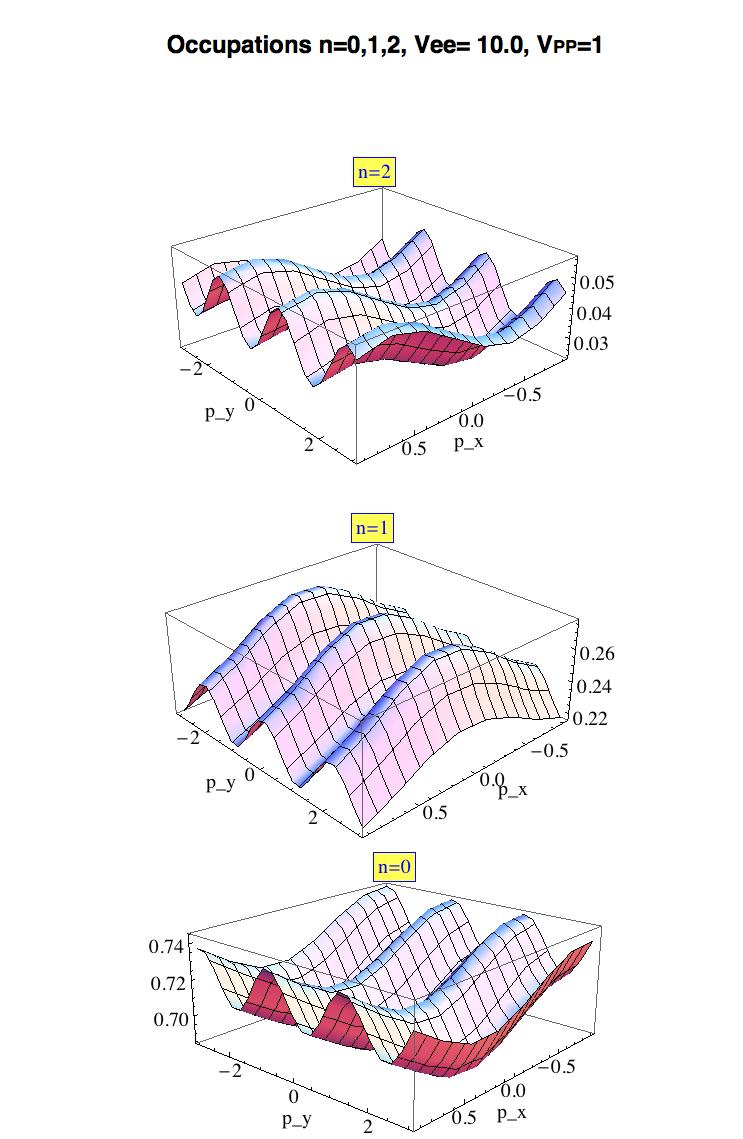}
\caption{The occupation numbers in CF-LLs 0,1,and 2 at
  $V_{ee}=10$,$V_{\mbox{\tiny PP}}=1$, $\omega=10$. Notice that the $n=0,1$ levels (bottom two) saturate the
  occupancy while the $n=2$ (top) level is practically empty,
  validating the truncation at three CF-LLs and establishing CF theory
  as a good low-energy description.  The plots repeat three times in
  the $p_y$ direction because $T_x$ and $T^{2}_{x}$ commute with $H$, but do not
  commute with $T_y$ and add ${2\pi \over 3}$ and ${4\pi \over 3}$ to
  $p_y$, respectively.\label{occ3}}
\end{figure}

\begin{figure}[t]
\includegraphics[width=8cm]{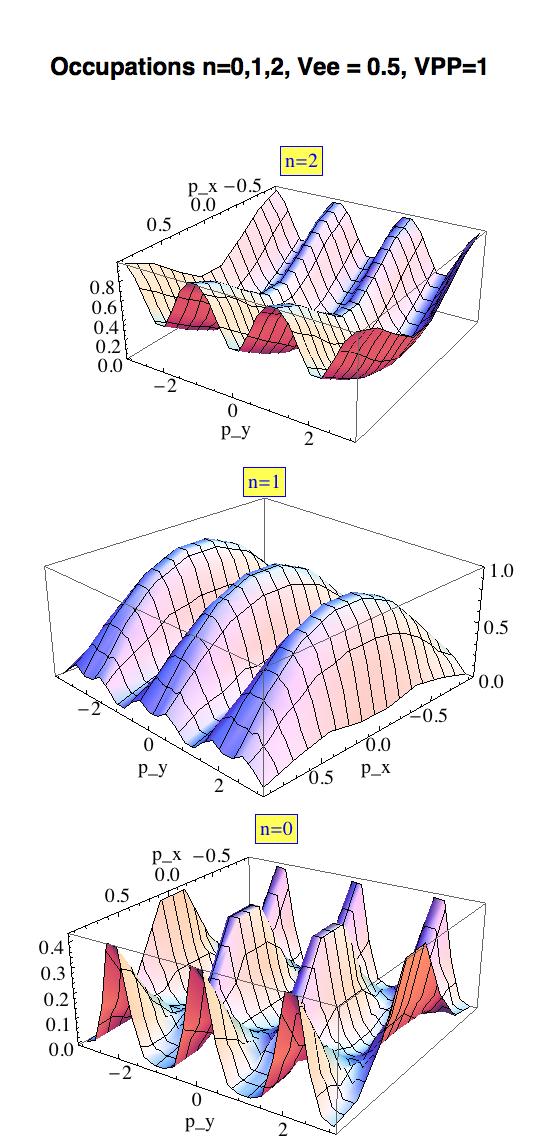}
\caption{The occupation numbers in CF-LLs 0,1,and 2 at
  $V_{ee}=0.5$, $V_{\mbox{\tiny PP}}=1$, $\omega=10$. Notice that the particles are levitating towards the
  $n=2$ level (top) and moving away from $n=0$, which calls the
  truncation at three CF-LLs into question.  The plots repeat three
  times in the $p_y$ direction because $T_x$ and $T^{2}_{x}$ which do
  not commute with $T_y$ add ${2\pi \over 3}$ and ${4\pi \over 3}$ to
  $p_y$, respectively, commute with $H$.  }
\label{occ05}
\end{figure}

With the Hamiltonian expressed in terms of CF operators, we move to
the HF calculation. We set $\omega =10$ (the gap between the two
electronic LLs ) and choose the periodic potential to be
$V_{\mbox{{\tiny PP}}}=V_{10}e^{-\half \pi}=1$. We keep 3 CF-LLs which
get mixed by the periodic potential and interaction. We end up with
three bands which are fairly well separated.  Unlike in the continuum
where the CF-LLs were uniformly filled, the occupation number here
varies with $\bp$ (due to the periodic potential) and had to be found
self-consistently.  Figure \ref{hf} shows the results of our
calculation for the coulomb interaction of strength ${2\pi V_{ee}
  \over q}$.  We see a clear gap separating the lowest band which is 
fully occupied from the others, even at very small values of $V_{ee}$.

This is puzzling since one expects the CF-picture to break down for
$V_{ee}\ll V_{\mbox{{\tiny PP}}}$.  Upon further investigation we
found two signals that point to the breakdown of the CF picture, one
internal and one external.
 
 The internal one involves the  occupation
numbers of the CF-LLs  in the ground state at each $\bp$. Figure
\ref{occ3} shows that at $V_{ee}=10$, $n_{{\tiny{CF}}}=0,1$ are
robustly occupied while $n_{{\tiny{CF}}}=2$ has negligible
occupancy. Thus our truncation with three CF-LLs is safe, since the
$n=2$ level is not called into play.  By contrast at $V_{ee}=0.5$ we
see in Figure \ref{occ05} that occupation of
$n_{{\tiny{CF}}}=2$ can be substantial.  The levitation of the
fermions to the upper CF-LLs is a clear indication that CF theory is
failing as a good low-energy theory.

Before moving to the external signature, let us observe a three-fold
symmetry in the occupations as a function of $p_y$, which is also
reflected in the energy bands. This is a consequence of the
$x$-translation symmetry of the noninteracting
Hamiltonian\cite{ming}. Recall that $T_x$ commutes with the
Hamiltonian, but not with $T_y$. It can be easily shown that

\beq
T_x|p_x,p_y,n\rangle=e^{i\phi(\bp)}|p_x,p_y-\frac{2\pi}{3},n\rangle
\eeq

This symmetry is also possessed by the HF Hamiltonian
\beq
T_x^{\dagger}H_{HF}(p_x,p_y)T_x=H_{HF}(p_x,p_y-\frac{2\pi}{3})
\eeq

and thus by the energy bands and the occupations.

\begin{figure}[t]
\begin{center}
\includegraphics[width=8cm]{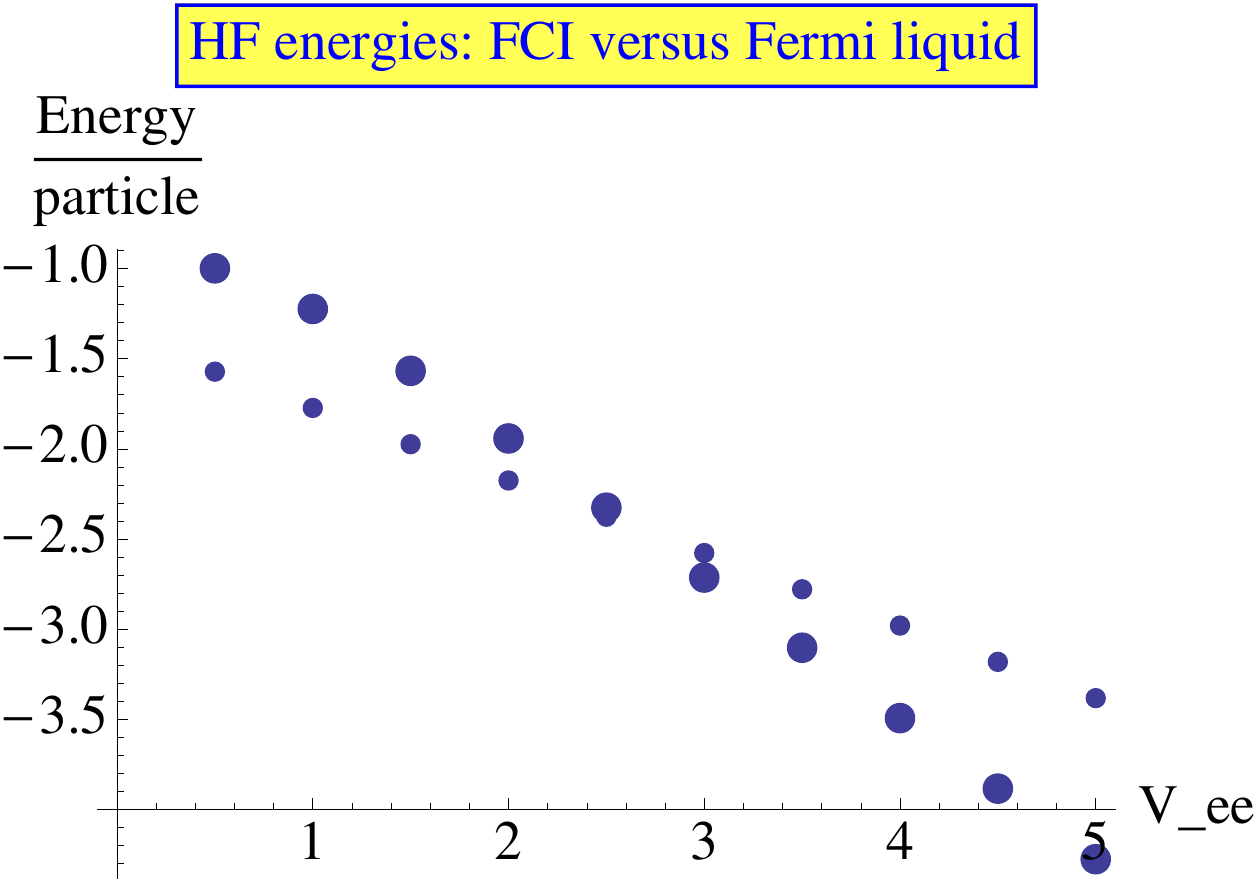}
\caption{Comparison of the HF energy per particle of the electronic
  Fermi liquid (smaller dots) versus the ${1\over 3}$ FCI state (larger
  dots). We see that the Fermi liquid, which wins at small interaction
  strength $V_{ee}$ yields to the FCI state around $V_{ee}=2.5$.}
\label{flvscf} 
\end{center}	
\end{figure}

Now we turn to the external test that signals the breakdown of the CF
state and is sharper than the one based on occupation numbers.  It
involves the comparison of the variational energy per particle in the
FCI state and the electronic Fermi liquid state.  It is clear that the
energy of the elecronic Fermi liquid state (which lies in the physical
Hilbert space) provides a variational upper bound for the exact ground
state energy.  It is not so clear that the energy of the CF state,
which is defined in a bigger space containing the physical coordinate
$\bR_e$ and the unphysical vortex coordinates $\bR_v$, is variational.
One might worry that these extra unphysical degrees of freedom may
allow a further lowering of energy not permitted in the physical
sector. This is, however, not true. Since the hamiltonian in the
enlarged space depends only on the electronic guiding center
coordinate $\bR_e$ via $\rho_{{\mbox {\tiny GMP}}}$ and is independent
of the vortex coordinate $\bR_v$, the exact eigenfunctions must be
tensor products of the exact eigenfunctions in the physical sector and
arbitrary wavefunctions in the unphysical sector.  However, the energy
is independent of the choice made in the unphysical sector. Thus the
exact ground state energy in the enlarged space is equal to that in
the physical sector and consequently {\em any function in the enlarged 
  space can furnish a variational upper bound to the exact ground
  state energy}. We compute the energy of the HF ground state with one
filled CF-LL. This, of course, is not an exact eigenfunction. In fact,
it is not even in the form of a tensor product between the physical
and unphysical sectors, being instead a linear combination of tensor product
states. However, by the above argument, it nevertheless provides a
variational upper bound on the exact ground state energy.
Fig. \ref{flvscf} shows the the HF energy per particle of the Fermi
liquid (smaller dots) versus to ${1\over 3}$ FCI state (larger
dots). We see that the Fermi liquid yields to the FCI state at
$V_{ee}\simeq 2.5$. We caution the reader that this does not mean that
the FCI state unequivocally wins. It is possible that there are
correlated {\em electronic} states with an even lower energy that our
CF-state.

To summarize, we produced a non-trivial Chern band by transferring the
topology to the basis functions in $\bp$ space. These functions arose
from two electronic LLs mixed by a periodic potential. The projected
charge density at momentum $\bq$ was then written as a computable sum
over $\bG$ of GMP densities at $\bq + \bG$. The GMP densities were
treated in the Hamiltonian method by the replacement $\bR_e = \bR +
\etab c$. Finally a HF calculation was carried out using a gapped CF
ground state. We see that although a FCI state always exists,even for
very weak interactions $V_{ee}$, the occupations of the higher CF-LLs
becomes smaller with increasing $V_{ee}$, providing an internal signal
of its stability.  A sharper limit for the goodness of the FCI state
is provided by the comparison to the variational energy of the
electronic Fermi liquid state, which  wins for $V_{ee}<2.5$, but gets bested by the FCI state for larger $V_{ee}$.

\subsection{Variants and limitations of Type I models }
A natural extension of the above example is to impose more complicated
periodic potentials to get more complicated $\bB (\bp)$s. Earlier in
our exploration\cite{ms1} we pursued this line of thought so as to reproduce the
$\bB (\bp)$ of some specific lattice model, say the LDM. (We see in
Fig. \ref{actual} that the $\bB (\bp) $ of the MLLL with just one
harmonic is already not too different from that of the LDM.) Our
motivation was as follows.  {\em Let us take the view that the FCB
  problem is defined by (i) $\bB (\bp) $ and (ii) the interaction
  written in terms of the projected density ${\rho}_{{\mbox{\tiny
        FCB}}}(\bq )$.}  The logic behind (i)
is that the projected electron coordinate
\begin{equation}
\big(R_{\mu}^{FCB}\big)_{\bp\bp'} = \big(i\frac{\partial} {\partial
  p_{\mu}} + \bA_{\mu}(\bp)\big)\delta^2(\bp-\bp')
\end{equation}
has commutators defined by $\bB$:

\beq
\left[R_{x}^{FCB},R_{y}^{FCB}\right]_{\bp\bp'} = i{\bB}(\bp)\delta^2(\bp-\bp')
\eeq

It follows that if we can construct a surrogate band with the same
$\bB (\bp )$ as in a given FCB,  {\em then any function of the
  projected electron coordinates will be algebraically the same in the
  two problems}. 

Let us review where this line of thought leads in the continuum FQH
problem. There we represent the electron's projected coordinate
$\bR_e$ in terms of CF variables $\bR$ and $\etab$. Getting the
algebra of $\bR_e$ correctly also means getting the algebra of
$\rho_{{\mbox{\tiny LLL}}}$, the projected density right, up to a
known prefactor.  Thus, if $\langle \cdots \rangle$ denotes averages
in the target band, which here is the LLL,\beqr \br_e &=& \bR_e +
\etab_e\ \ \ \mbox{which implies} \\ \langle \br_e\rangle &=&
\bR_e\ \ \ \ \ \mbox{ while} \\ \langle e^{i \bq \cdot \br_e} \rangle
&=& \langle e^{i \bq \cdot (\bR_e + \etab_e)} \rangle\\ &=& \langle
e^{i \bq \cdot \etab_e} \rangle e^{i \bq \cdot \bR_e} \\ &=&
e^{-q^2l^2/4} e^{i \bq \cdot \langle \br_e \rangle} \eeqr

In other words, the projection of the exponential of $\br_e$ is, up to
a known prefactor $e^{-q^2l^2/4}$, the same as the exponential of the
projection because $\etab_e$ and $\bR_e$ commute. This is why if the 
projected coordinate is faithfully represented, so is the projected
density.  We go over these well known facts to highlight the unusual
simplicity of projecting to the LLL.

Unfortunately, in the generic FCB problem this is no longer true. Let
us define two different projected densities in the FCB. One is the
usual one: 

\beq {\rho}_{{\mbox{\tiny FCB}}}(\bq) = \langle \mbox{FCB}| e^{i\bq
  \cdot \br_e}|\mbox{FCB}\rangle \eeq This is the projected density
which enters the interacting Hamiltonian in the FCB.

The other is the analogue of the guiding center density:

\beq
\bar{\rho}_{{\mbox{\tiny FCB}}}(\bq )=e^{i\bq\cdot\bR_e^{\mbox{\tiny FCB}}} 
\eeq

In the FCB problem

\beq {\rho}_{{\mbox{\tiny FCB}}}(\bq)\ne C(q) \bar{\rho}_{{\mbox{\tiny
      FCB}}}(\bq ) \eeq 

because the ``guiding center'' coordinates $\bR^{\mbox{\tiny FCB}}$ do
not commute with the analogue of the cyclotron coordinates. {\em Thus,
  $\bofp$ is not enough to specify the interacting Hamiltonian in the
  FCB completely. One needs the expression for ${\rho}_{{\mbox{\tiny
        FCB}}}(\bq)$ as well}.
  
However, to   lowest order in  $\bq$ and $\bq'$
 we have, in first quantization and in $\bp$-space,
\beq
\left[ {\rho}_{{\mbox{\tiny FCB}}}(\bq), {\rho}_{{\mbox{\tiny FCB}}}(\bq')\right] = -i \left[ \bq \times \bq' \right] \bB (\bp ) + \mbox{higher order terms}
\eeq
as pointed out by Parameswaran {\em at al} \cite{sid}.

So if the Chern flux density $\bB (\bp)$ of the surrogate band matches
that of the lattice FCB, ${\rho}_{{\mbox{\tiny FCB}}}(\bq)$ and
$\bar{\rho}_{{\mbox{\tiny FCB}}}(\bq )$ will bear a close resemblance
to each other but not be equal in all respects.  Since these are both
models anyway, one may argue that it is sufficient to get a surrogate
that approximates the original lattice model and is yet amenable to
analytic treatment. However we will not pursue this approach further
since there is a more direct way to obtain ${\rho}_{{\mbox{\tiny
      FCB}}}(\bq)$ in terms of $\rho_{\mbox{\tiny GMP}}(\bq)$ for an
arbitrary lattice model, which we now describe.

\section{Type II models}

In the previous section we showed how to carry out the CF substitution
in LL-based models with nonconstant $\bofp$. In this section we present a more
general approach for incorporating CFs in the solution of any {\em given}
lattice model with a Chern band, without any reference to LLs. We illustrate this approach by considering the LDM with some
interaction $V_{ee}$.

Here are the concrete set of steps we follow:
\begin{itemize}
\item Construct $\rho_{{\mbox {\tiny FCB}}}(\bq)$, the projected
  density operator in terms of the eigenfunctions of noninteracting
  LDM hamiltonian $H(\bp )$ and the operators $d$ and $d^{\dag}$
  associated with the Chern band at each $\bp$.
\item Construct operators obeying the algebra of $\rho_{{\mbox {\tiny
      GMP}}}(\bq +\bG)$ using $d$ and $d^{\dag}$ as per
  Eq. \ref{rhoq+g}. As noted immediately after that equation, this can
  always be done. To get the best possible results this must be done
  in $A_y=0$ gauge.
\item Expand as before, using the complete set of $N^4$ operators
         $\rho_{{\mbox {\tiny GMP}}}(\bq +\bG)$:
        \beq
        \rho_{{\mbox {\tiny FCB}}}(\bq)=\sum_{\bG}c(\bG, \bq) \rho_{{\mbox {\tiny GMP}}}(\bq +\bG).
        \eeq
This will be just a Fourier expansion in $e^{ip_xn_y-ip_yn_x}$.
\item Carry out the CF substitution in $\rho_{{\mbox {\tiny GMP}}}$ and go on to the HF approximation.
\end{itemize}

We illustrate  the above steps with the LDM at $M=1$:
    \beq
   H (\bp) = \sigma_1 \sin p_x + \sigma_2 \sin p_y +\sigma_3 (1 - \cos p_x- \cos p_y) \label{ldh2}
   \eeq 

with ground-state energy

   \beq
   -\e (\bp)=-\sqrt{1 + 2 (1- \cos p_x)(1-\cos p_y)}.
   \eeq

   The generic formula for the projected charge density is 

   \beq
        {\rho}_{{\mbox{\tiny FCB}}}(\bq) =\sum_{\bp}  d^{\dag}(\left[ \bp + \bq \right]) d(\bp) \langle \left[ \bp + \bq \right]|  \bp \rangle \label{formal}
    \eeq

where $d^{\dag}\ , d$ create and destroy the ground state and
$|\bp\rangle$ is the corresponding Bloch spinor. However $|\bp\rangle$
is not globally defined\cite{thouless2} over the BZ because ${\cal C}=-1$.  Here are
two choices that work in two different patches:

   \beq
   | \bp \rangle_{1}^{0} = \left(
           \begin{array}{c}
             \sin {\theta (\bp )\over 2} \\ - \cos {\theta (\bp ) \over 2}e^{i \phi (\bp) }
           \end{array}
         \right)
\eeq

  \beq
   | \bp \rangle_{2}^{0} = \left(
           \begin{array}{c}
             \sin {\theta (\bp )\over 2} e^{-i \phi (\bp) }\\ - \cos {\theta (\bp ) \over 2}
           \end{array}
         \right)
\eeq

where
\beqr
\cos \theta (\bp) &=& {1 - \cos p_x-\cos p_y \over \e (\bp )}\label{cos}\\
e^{i \phi (\bp )}&=& {\sin p_x + i \sin p_y \over \sqrt{\sin^2 p_x + \sin^2 p_y}}\label{expphi}
\eeqr

The superscript $0$ on the kets reminds us that these are going to be
transformed to a more appropriate gauge later.

\begin{figure}[b]
\includegraphics[width=8cm]{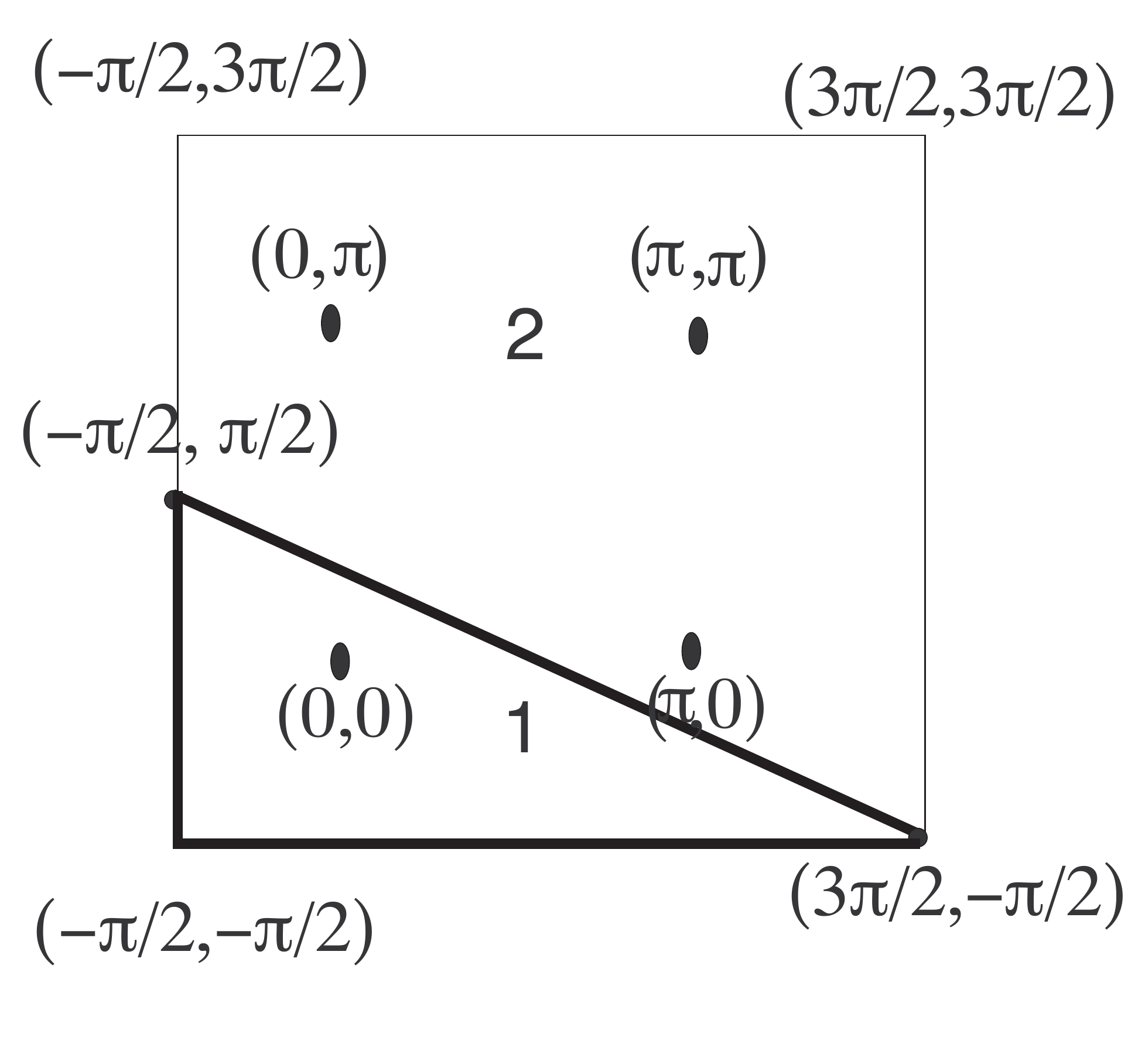}
\caption{In patch $1$, we have the south pole at $(0,0)$ and spinor
  $|\bp\rangle_1$ is well defined while in patch $2$, spinor
  $|\bp\rangle_2$ is well defined at the north pole reached at the
  points $(0,\pi), (\pi , 0), (\pi, \pi)$. The patches meet on the
  right triangle. In the final ${\cal A}_y=0$ gauge the wavefunction is
  seamless across the hypotenuse $p_y = {\pi \over 4}-{p_x\over 2}$
  and periodic in $p_x$.  However the top and bottom edges differ by a
  phase and the Chern number is resident in that difference. }
\label{patchez}
\end{figure}

The angle $\phi (\bp) $  is ill defined at $\theta (\bp) = 0 , \pi$. At $\theta (\bp) = 0$, choice $| \bp \rangle_2$ is good because the component  $\sin {\theta (\bp )\over 2} e^{-i \phi (\bp) }$ vanishes. In a patch where $\theta (\bp)=\pi$, the choice $| \bp \rangle_1$ is good. Figure \ref{patchez} shows that there are four trouble spots: $(0,0)$ where the spinor is at  the south pole $\theta = \pi$,  and points $(0,\pi), (\pi, \pi)$ and $(\pi , 0)$ where it is at the north pole $\theta =0$. We pick the BZ in the range $\left[ -{ \pi \over 2},  {3\pi \over 2}\right]$ so that the trouble  spots are not at the edges. Patch $1$ is indicated by the solid right triangle and patch $2$ is the rest.

 The corresponding Berry connections are
\beqr
\bA_{1  }^{0}  (\bp)&=&- \half (1 + \cos \theta (\bp))\nablab \phi\\
\bA_{2  }^{0}  (\bp)&=& \half (1 - \cos \theta (\bp))\nablab \phi
\eeqr

Once again the superscript $0$ on the $\bA^0$'s above signals that
this is not yet the final gauge.

If we imagine the edges of the BZ parallel to $\be_y$ glued together
to form a cylinder and then the top and bottom sewn together to form the
torus, this dark line will mark the boundary between the two regions.
The difference between the two connections is $-\nablab \phi$, and the
integral around the boundary of $-{1 \over 2\pi}\nablab \phi$ will
yield $ {\cal C}=-1$.

 We now  reach the final gauge
${\cal A}_y=0$  as follows.  Let us define
 \beqr
 \Lambda_1(p_x,p_y)&=& \int_{-\half \pi}^{p_y}{\cal A}_{1y}^{0}(p_x,p_{y}{'})dp_{y}^{'}\\
 \Lambda_2(p_x,p_y)&=& \Lambda_1 (p_x, {\pi \over 4}-{p_x\over 2})+\int_{{\pi \over 4}-{p_x\over 2}}^{p_y}{\cal A}_{2y}^{0}(p_x,p_{y}{'})dp_{y}^{'}\nonumber\\
 & & \\
 \chi (p_x)&=&
 \phi (p_x, {\pi \over 4}-{p_x\over 2})
 \eeqr
 and the following final kets in the two patches
 \beqr
 |\bp\rangle _1 &=& e^{i\Lambda_1}|\bp\rangle _{1}^{0}\label{patch1}\\
 |\bp\rangle _2 &=& e^{i\Lambda_2 +i\chi}|\bp\rangle _{2}^{0}\label{patch2}
 \eeqr

In this final gauge, not only is ${\cal A}_y=0$, the two expressions above,
$|\bp\rangle _1$ and $|\bp\rangle _2$, merge seamlessly along the
sloped line $p_y={\pi \over 4}-{p_x\over 2}$ separating the patches
and on the vertical boundaries of the BZ, which may be glued to form a
cylinder. However between  the lines $p_y=-{\pi \over 2}$ and $p_y={3\pi
  \over 2}$ there is a phase mismatch so that we cannot roll the
cylindrical BZ to a torus without a discontinuity at the seam.  This
had to be so, for otherwise we would have a fully periodic Bloch
function and ${\cal C}$ would vanish by Stokes' Theorem \cite{thouless2}.

A salient feature of this gauge is that ${\rho}_{{\mbox{\tiny
      FCB}}}(\bq)$ in Eq. \ref{formal} also has jumps in the sum over
$\bp$ exactly where the ${\rho}_{{\mbox{\tiny GMP}}}(\bq +\bG)$'s do,
rendering this the optimal gauge for rapid convergence of the Fourier
expansion.
 
\begin{figure}
\includegraphics[width=8cm]{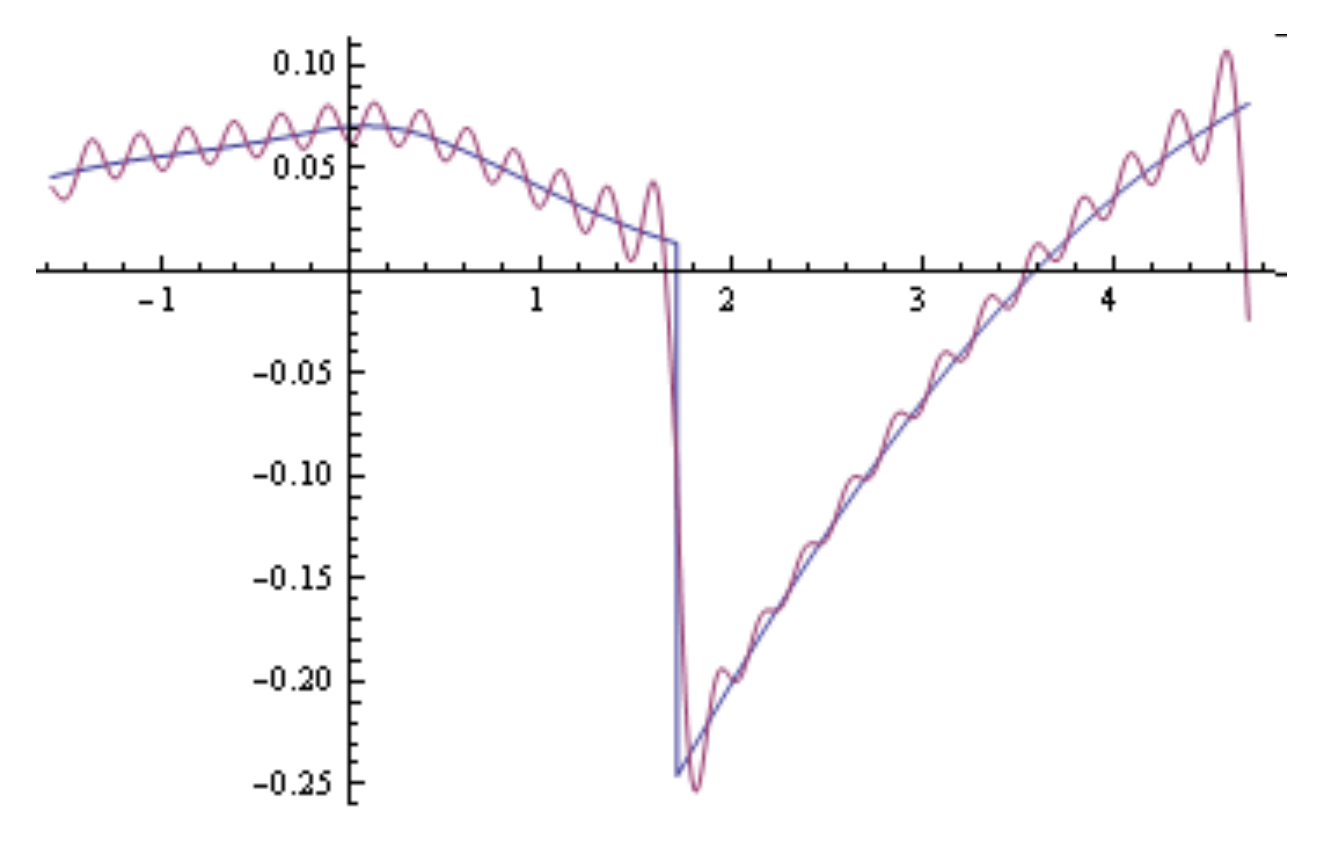}
\caption{Here we compare, at fixed $p_x=-{\pi \over 2}+5$, the
  imaginary part of the coefficient of $d^{\dag}(\left[ \bp +
    \bq\right] )d(\bp)$ in ${\rho}_{{\mbox{\tiny FCB}}}(\bq)$ to the
  approximation in which the Fourier sum over $\bG$ is truncated after
  $50$ harmonics in the $p_y$ direction (from $-25 \ \mbox{to } \ 25$)
  and $20$ in the $p_x$ direction.}
\label{imrocompfixedy}
\end{figure}

Figure \ref{imrocompfixedy} gives a taste of how the Fourier expansion
works at a generic value of $\bq = 3 \be_x+ 3 \be_y$. It compares, at
fixed $p_y=-{\pi \over 2}+5$, the imaginary part of the coefficient of
$d^{\dag}(\left[ \bp + \bq\right] )d(\bp)$ in ${\rho}_{{\mbox{\tiny
      FCB}}}(\bq)$ to the approximation in which the Fourier sum over
$\bG$ is truncated after $50$ harmonics (from $-25 \ \mbox{to } \ 25$)
 in the $p_y$ direction and $20$ in the $p_x$ direction. Note
that even though $\bq$ has components in both directions, the jump
occurs only in the $p_y$ direction due to the periodicity in the $p_x$
direction.

\begin{figure}[h]
\includegraphics[width=8cm]{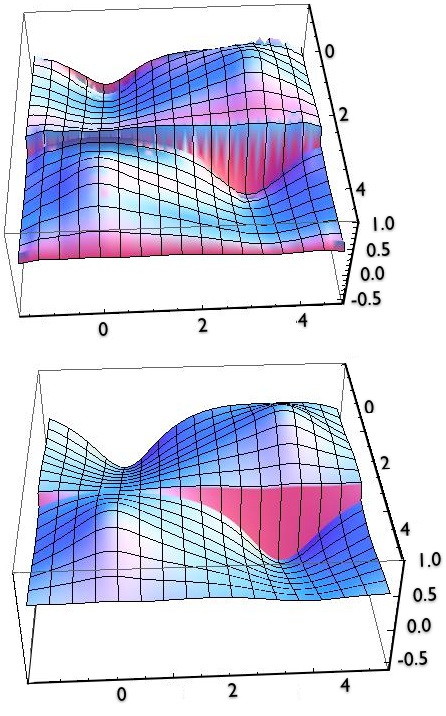}
\caption{The upper half shows the full landscape of the real part of
  the coefficient of $d^{\dag}(\left[ \bp + \bq\right] )d(\bp)$ in
  ${\rho}_{{\mbox{\tiny FCB}}}(\bq)$ in the approximation with $50$
  harmonics (from $-25\mbox{ to } +25$) in the $p_y$ direction and
  $20$ in the $p_x$ direction. The lower one shows the actual values.
  }
\label{rerocomp}
\end{figure}

Figure \ref{rerocomp} shows the full landscape of the the real part of
the coefficient of $d^{\dag}(\left[ \bp + \bq\right] )d(\bp)$ in
${\rho}_{{\mbox{\tiny FCB}}}(\bq)$ versus the approximation in which
$50$ harmonics (from $-25 \ \mbox{to } \ 25$) are kept in the $p_y$
direction and $20$ in the $p_x$ direction.

Having expressed everything in terms of ${\rho}_{{\mbox{\tiny
      GMP}}}(\bq)$, the the CF substitution and HF analysis can be
carried out just as before and we do not discuss it further.

A central message of this work is that since in a problem with ${\cal
  C}\ne 0$, one cannot work with periodic Bloch functions, the
expression for ${\rho}_{{\mbox{\tiny FCB}}}(\bq)$ in the ${\cal
  A}_y=0$ gauge will necessarily have a jump beyond some $p_y$
depending on $q_y$ when we retract from a point $\bp +\bq$ which lies
outside the BZ to a point $\bp'=\left[ \bp+ \bq\right]$ within. The
GMP density ${\rho}_{{\mbox{\tiny GMP}}}(\bq)$, has exactly such a
jump and is the right basis to use. When the Chern band supports an
FCI, CF coordinates are the natural variables in terms of which the system can be understood in the simplest way.

 \section{The case of trivial bands}
Let us ask if we have achieved ``too much''. Consider a topologically
trivial band with Chern number zero. Nothing prevents us from
representing its projected density in terms of ${\rho}_{{\mbox{\tiny
      GMP}}}(\bq + \bG)$

\beq \bar{\rho}
(\bq) = \sum_{\bG} c(\bG) {\rho}_{{\mbox{\tiny GMP}}}(\bq + \bG),
\eeq

and carrying out the CF substitution. So, are there FCIs in
topologically trivial bands?

\begin{figure}[h]
\includegraphics[width=9cm]{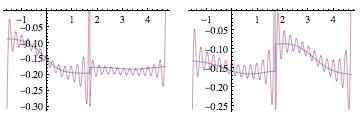}
\caption{Comparison, in the case with ${\cal C}=0$,  of imaginary parts of exact and approximate densities at $p_x= -{\pi \over 2}+1$ and $p_x=-{\pi \over 2}+6$ with the approximation truncated after $50$ harmonics ($+25 $ to $-25$) in the $p_y$ direction}
\label{triv}
\end{figure}

This appears to be a subtle issue. On the face of it, since the Bloch
spinor can now be globally defined as a periodic function in the BZ,
so can ${\rho}_{{\mbox{\tiny FCB}}}$, and its expansion in terms of
${\rho}_{{\mbox{\tiny GMP}}}$ which has a jump in the BZ seems ill
fated. Although completeness assures us that with infinite number of
terms we can do it, in order to give the expansion the best chance, we
must first transform the spinor to the gauge ${\cal A}_y=0$, just like the
functions entering ${\rho}_{{\mbox{\tiny GMP}}}$. This makes
${\cal A}_x(p_x,p_y+2\pi)\ne {\cal A}_x(p_x,p_y)$ and causes the familiar jump.

The jump of this sort is inevitable if $B(\bp )$ is non-constant,
because in this gauge 

\beq {\cal A}_x(p_x,p_y)= -\int_{-{\pi \over
    2}}^{p_y}\bB (p_x,p_{y}^{'} )dp_{y}^{'} \eeq 

and this integral
need not vanish at any fixed $p_x$. 
 
The Fourier expansion, while not so successful as in the case ${\cal
  C}\ne 0$, is still promising, as shown see Fig. \ref{triv} for two
slices at $p_x= -{\pi \over 2}+1$ and $p_x=-{\pi \over 2}+6$.  The key
feature is that the smaller the jump (i.e. the smaller the magnitude
of $\bofp$) the more Fourier components it takes to approximate it to
a given accuracy.

The bottom line is that, under certain conditions, even a band with
${\cal C}=0$ could exhibit FQHE under partial filling, an issue we are
actively pursuing. 

So far we have ben able to rule out the following completely trivial
case: a band in which $\bB (\bp)\equiv 0$ with no band dispersion. The
energy is all potential and proportional to $V_{ee}$. Although an FCI
state exists at $\nu = { 1\over 3}$, its energy is always higher (by a
factor of roughly $2$) than the energy of the electronic Fermi liquid.
Thus, in this case at least, FQHE does not appear to be favoured in
the topologically trivial band.

\section{Fractions where $\nu\ne \sigma_{xy}$}
\begin{figure}[t]
\begin{center}
\includegraphics[width=8cm]{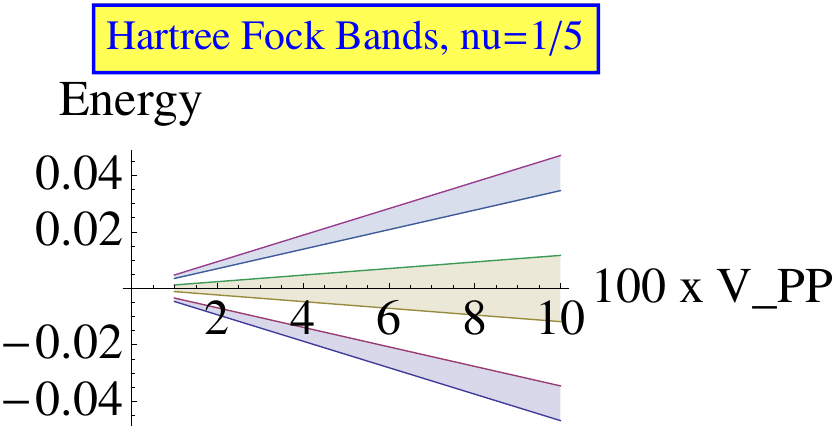}
\caption{The highest and lowest energies of the three sub-bands for
  the exotic fraction ${1 \over 5}$ as a function of the parameter
  $V_{PP}$.  The bandwidths grows with
  the strength of the periodic potential but are still less than the sub-band
  gaps. The other parameters are $V_{ee} = .1, \omega
  =1$.}
\label{hfkr} 
\end{center}	
\end{figure}
\begin{figure}[t]
\begin{center}
\includegraphics[width=8cm]{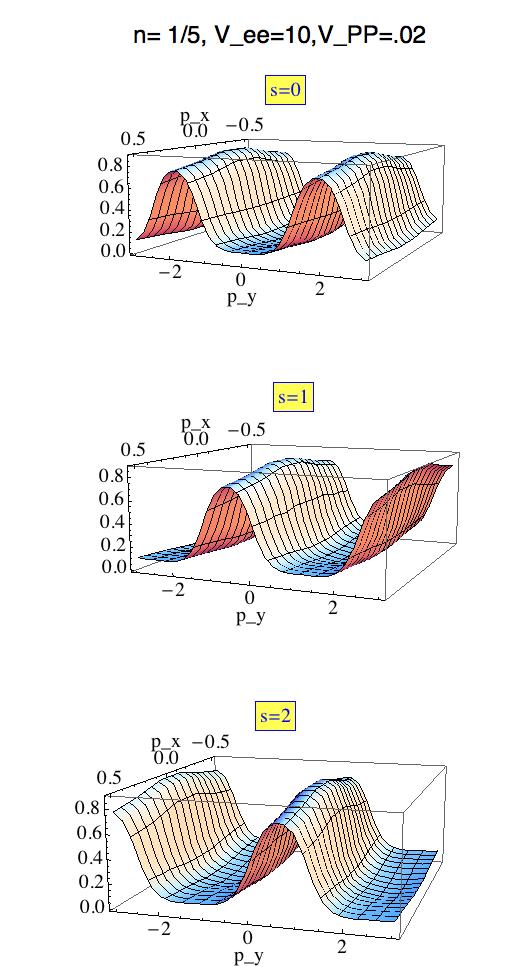}
\caption{From the top, the occupation numbers of the three CF-LLL- sub-bands, s=0,1,2,  for the exotic fraction $\nu={1 \over 5}$. The action of $T_{x}^{1}, T_{x}^{2},T_{x}^{3},T_{x}^{4}$ is to  change $p_y$ by multiples of 
${6\pi \over 5}$ and change the sub-band index by $1$ whenever $p_y$ changes by $2\pi$. If the pictures of the three sub-bands are glued to together in the order $0,1, 2$, we see $5$ oscillations altogether.  \label{occupkr} }
\end{center}	
\end{figure}

In the presence of a periodic potential, the equality $\nu=
\sigma_{xy}$ is not mandatory, since Galilean invariance is explicitly
broken.  Exotic states whose very existence depends on an interplay of
interactions and the lattice potential are
possible\cite{kol-read}. There is suggestive numerical evidence for such states in a
problem of lattice hard-core bosons in an external
field\cite{sorensen,moller}. 

Since we have expressed the FCB Hamiltonian in CF language, it is
natural for such states to be realized in FCBs under suitable
conditions.

Consider a case with $\nu = {1 \over 5}$. Each electron sees $5$ flux
quanta. Let us attach $2$ (and not the usual $4$) flux quanta to each
electron to form CFs. Each CF sees $3$ flux quanta and
$\nu_{\mbox{{\tiny CF}}}={1 \over 3}$. In the continuum this partially
filled CF-LLL is gapless, and thus not stable. However, as we will now
show, such a state is generically gapped in the presence of a periodic
potential. Since $e^*/e=3/5$, the CF sees ${3 \over 5}$ flux quanta in
the electronic unit cell. So we must take $5$ times the electronic
cell (say in the x-direction) to form a CF unit cell that encloses
integer flux ($3$) and ensures that the CF translation operators
$T_{x}^{5}$ and $T_y$ commute. In general if there are $p/q$ quanta of
effective flux per unit cell, each CF-LL will split into $p$
sub-bands\cite{TKNN}. Thus, in our example each CF-LL will spilt into three
sub-bands. If the lowest of these three sub-bands, when filled, is
separated from the others by a gap, we obtain an FCI.
 
Our numerical work fully corroborates this picture. We work to linear
order in the periodic potential $V_{PP}$, and keep only the lowest
CF-LL. Note that the treatment is not perturbative in the interaction
strength. Fig. \ref{hfkr} shows
the three sub-bands whose widths grow with the strength of the
periodic potential while the gaps grow even faster. Figure
\ref{occupkr} is very interesting.  Once again $T_{x}^{1},
T_{x}^{2},T_{x}^{3},T_{x}^{4}$ commute with $H$ but not $T_y$. Each
power of $T_x$ changes $p_y$ by ${6\pi \over 5}$. Starting with $p_y
=-\pi$ of the sub-band $s=0$, if we act with these powers of
$T_x$, we move to $s=1$ when the momentum change is $>2\pi$, and to
$s=2$ when the momentum change is $>4\pi$.  If the three pictures are
glued end to end, we see $5$ full oscillations. (Compare this to the
$\nu ={1\over 3}$ case where three periods occurred within the same
CF-LL whereas here the $5$ oscillations are spread over $3$
sub-bands.)
  
What will be the Hall conductance of this state? We know from Kol and
Read \cite{kol-read} that  (in our convention where ${\cal C}=-1$ for a filled LL)

\beq \sigma_{xy}= {\sigma_{{\mbox{\tiny
        CF}}}\over 1 +2 \sigma_{{\mbox{\tiny CF}}}}\eeq 

where $\sigma_{{\mbox{\tiny CF}}}$, the dimensionless CF Hall
conductance (equal to minus its Chern number in our convention) of the
filled sub-band could be any integer. (When a LL splits into sub-bands
we only know that the sum of the Chern numbers of the sub-bands equals
that of the original LL. ) But no matter what this integer is, the
possible values of $ \sigma_{xy}= {1\over 3}, {2 \over 5}, {3 \over
  7}..$ do not include ${1 \over 5}$. The actual value happens to be
${2\over 5}$ because $\sigma_{{\mbox{\tiny CF}}}=2$.
 
\section{Summary and outlook}

We have demonstrated here that the Hamiltonian theory of the FQHE
\cite{rmp}, which was very useful in describing the FQHE in the
continuum, is just as effective in describing fractionally filled
Chern bands that exhibit FQH-like effects. This is surprising in view
of the fact that since there is no external magnetic field in the FCB,
ideas of flux attachment seem doomed from the outset.  This is not an
issue in our Hamiltonian theory which relies on an exact algebraic
mapping that expresses the projected density of the Chern band
$\rho_{\mbox{\tiny FCB}}(\bq)$ in terms of CFs in two steps. First, we
express $\rho_{\mbox{\tiny FCB}}$ as a linear combination of operators
$\rho_{\mbox{\tiny GMP}}$ satisfying the magnetic translation
algebra\cite{GMP}.  Second, we perform the CF substitution in
$\rho_{\mbox{\tiny GMP}}$ exactly as we did in the continuum
theory\cite{rmp}. These mappings of operator algebras can be carried
out for arbitrary lattice models with no reference to Landau levels.

We have illustrated the power of the approach by solving two
models. The first model is a $\nu=\third$ FCI and is adiabatically
connected to the continuum Laughlin state at the same filling.  The
second is a more exotic, $\nu=\fifth$ FCI relying on the lattice
potential for its very existence\cite{kol-read}. Its dimensionless
Hall conductance is ${2\over 5}$ and not ${1\over 5}$ as would be
expected in Galilean invariant state.  In both cases we have computed
the band structure of CFs in the Hartree-Fock approximation.

There are many interesting directions which we intend to pursue in
future work. Collective excitations for fractionally filled Chern
bands can be computed in a conserving approximation in our
approach\cite{read,pasqier,rmp}, as can finite temperature effects\cite{rmp}.

A specially interesting case is at $\nu =\half$ in a FCB. One expects
an electronic Fermi liquid at weak coupling and an HLR-type CF-Fermi
liquid at strong coupling. This transition, which we propose to study
by our operator-based method, has already been explored in the parton
formulation recently\cite{mcgreevy}.

Let us turn to quasiparticle excitations.  In addition to
Laughlin-type quasiparticles with fractional charge and statistics,
the lattice allows us to consider excitations not conceivable in the
continuum, such as those associated with lattice vacancies or
dislocations\cite{barkeshli-qi}.

As stated in the introduction, fractionally filled 2D time-reversal
invariant TIs\cite{levin,tang,sun,neupert,sheng,bern,sheng-wang,bern2}
can be treated by labelling the pair of Chern bands making up the
noninteracting TI (with equal and opposite Chern index) by a
pseudospin index. There is no requirement of $S_z$ conservation, and
the interactions can produce states which can spontaneously break
time-reversal, and/or states of the Kol-Read type. In fact, in a
numerical diagonalization on a small system Neupert {\it et
  al}\cite{neupert2} find a state in a regime of parameters which has
a filling of ${2\over 3}$ but a degeneracy of 3 (rather than the
degeneracy of $3\times3=9$ one would expect for ``independent''
$\nu=\third$ for each pseudospin) on the torus, suggesting that it
could be a Kol-Read\cite{kol-read} type state.

While finishing this manuscript we noticed the work of Grushin {\em et
  al}\cite{grush}, who have examined the conditions for the stability
of the FCI.

We thank the NSF for grants DMR-0703992 and PHY-0970069 (GM),
DMR-0103639 (RS), Yong-Baek Kim, Herb Fertig, Sid Parameswaran, and
Nick Read for illuminating discussions and Anne-Frances Miller for
help with the graphics. We are grateful to Shivaji Sondhi and Sid
Parameswaran for urging us to confront the numerics. Finally, we are
grateful to the Aspen Center for Physics (NSF 1066293) for its
hospitality and facilitating our collaboration.

\end{document}